\documentclass[prd,12pt,nofootinbib,tightenlines]{revtex4}

\usepackage{graphicx}
  \usepackage{amssymb}
   \usepackage{amsmath}
    \usepackage{amsfonts}
     \usepackage{epsfig}
      \usepackage{bm}

\begin{document}

\title{Complex mass definition and the concept of
continuous mass}

\author{V.~I.~Kuksa}
\email{kuksa@list.ru} \affiliation{Institute of Physics, Southern
Federal University, Rostov-on-Don, Russia}

\begin{abstract}
Propagators of unstable particles are considered in the spectral
representation which naturally follows from the concept of
continuous mass. The spectral functions are found with the help of
the most general formal and physical assumptions. Dressed
propagators of unstable scalar, vector, and spinor fields are
derived in an analytical way for a variant of parameter space.
 The structure of the propagators is in a
correspondence with the complex mass scheme.

\end{abstract}

\maketitle

\section{Introduction}
 Two standard definitions of the mass and
width of unstable particles (UP), which are usually considered in
the literature, have essentially different nature. The on-mass-shell
(OMS) scheme defines the mass $M$ and width $\Gamma$ of a bosonic UP
by the renormalization conditions \cite{1}:
\begin{equation}\label{1.1}
M^2=M^2_0 +\Re \Pi(M^2),\,\,\,M\Gamma=\frac{-\Im \Pi(M^2)}{1-\Re
\Pi^{'}_s(M^2)}\,.
\end{equation}
In the pole scheme (PS) the definitions of the mass and width are based
on the complex-valued position of the propagator pole
$s_R-M^2_0-\Pi(s_R)=0$. For instance, in the case of a vector UP a possible definition is as follows \cite{1}:
\begin{equation}\label{1.2}
s_R=M^2_{\rho}-iM_{\rho}\Gamma_{\rho},\,\mbox{where}\,\,s_R=M^2_0+\Pi(s_R).
\end{equation}
It should be noted that this definition of the pole mass $M_{\rho}$
and width $\Gamma_{\rho}$ (also called $m_2, \Gamma_2$ or
$\tilde{M}_Z, \tilde{\Gamma}_Z$) is not unique. There has been
considerable discussion concerning definition of the vector-boson
mass \cite{2,3,4,5,6,7,8}. It was shown that OMS scheme contains
spurious higher-order gauge-dependent terms while PS provides gauge
invariant definition. Later on the PS definition was considered in
detail \cite{9,10,11} and gauge-invariant treatment was developed in
the frame of the complex-mass scheme (CMS) \cite{12,13}.

An alternative, semi-phenomenological, scheme based on the
hypothesis of continuous (smeared) mass of UP was considered in
\cite{14,15} (and references therein). In this approach, the
physical values of the mass and width are related to the parameters
of continuous mass distribution. In the model the propagators of UP have
spectral-representation form, so that the standard problems
transform to constructing the spectral function which describes the
mass distribution. The effective theory of UP was developed on the
base of this approach in \cite{16}, where the method of
factorization of widths and cross-sections is presented.

In this work, we show that the concept of continuous mass along with
some natural assumptions (which are considered in the next section)
lead to the pole scheme with complex mass definition. This result
follows directly from the propagators in spectral
representation which, in turn, follows from the quantum-field
realization of the mass-smearing concept \cite{14,15}.

The paper is organized as follows. In section~\ref{sec:2} we present the
principal elements of the approach under consideration and analyze
the general structure of the propagators in the model. The structure of the
bosonic and fermionic propagators is considered in sections~\ref{sec:3} and \ref{sec:4}
respectively. Some conclusions concerning the physical status of the
results are made in section \ref{sec:5}.

\section{\label{sec:2}General structure of the propagator of an unstable particle}

Traditional way to construct the dressed propagator of UP is the
Dyson summation, which introduces the width and redefines the
mass of UP. This procedure runs into some problems widely discussed
in the literature. Let us consider one of the formal sources of the
difficulties which follow from this summation. In the simplest case
of a scalar UP this procedure looks like
\begin{align}\label{2.1}
D_{(1)}(q)={}&D_0(q)\sum_{k=0}^{\infty}(-i\Pi_{(1)}(q)D_0(q))^k\notag\\
={}&\frac{D_0(q)}{1+i\Pi_{(1)}(q)D_0(q)}=\frac{i}{q^2-M^2_0-\Pi_{(1)}(q)}\,,
\end{align}
where $D_0(q)=i(q^2-M^2_0+i\epsilon)^{-1}$ is the propagator of a scalar
free field and $\Pi_{(1)}(q)$ is the one-particle--irreducible
self-energy. The dressed propagator (\ref{2.1}) is formally
incorrect in the near-resonance range, because of the finite radius of convergence of the series
\begin{equation}\label{2.2}
\frac{1}{1-z}=\sum_{k=0}^{\infty}z^k=1+z+z^2+...,\qquad |z|<1.
\end{equation}
From Eqns.~(\ref{2.1}) and (\ref{2.2}) it follows that the variable
$z=\Pi_{(1)}(q)/(q^2-M^2_0)$ should be correctly redefined before
summation and one can not use the procedure in the region $z>1$
after the redefinition. Thus, the dressed propagator in the form
(\ref{2.1}) contains an assumption of infinite series, that is, UP
is described as a non-perturbative object.  This, evidently, leads
to difficulties in the scheme of sequential fixed-order
calculations, which exhibit themselves in the violation of gauge
invariance. As was pointed out in the previous section, such
problems have been under considerable discussion for many years.

An alternative approach is based on the spectral representation of
the propagator of UP. It has a long history \cite{17,18,19,20,21}
and treats UP as a non-perturbative state or effective field
(asymptotic free field \cite{20,21}). Here, we consider this
approach in the framework of the model of UP with continuous
(smeared) mass \cite{14,15}. In the works \cite{14,15}, the
effective field function of UP is a continuous superposition of
ordinary ones with a universal weight function $\omega(m)$ which
describes the distribution of the mass parameter $m$. The value $m$
was interpreted as a random mass with continuous distribution from
some threshold to infinity (physical range definition). Here, we
consider some modification of this approach which leads to
significant changes in the effective theory of UP. We consider a
non-universal weight function $\omega(q,m)$, that is the function
which depends on the momentum $q$ of the state. The second
modification is the extension of the allowed values of the mass
parameter $m^2$ from $[m^2_0,+\infty)$ to $(-\infty,+\infty)$. So,
the parameter $m^2$ loses its previous simple interpretation. It
acquires, however, the status of the momentum squared $q^2$ in the
whole Minkowski space. Further we show that this modifications lead
to PS definition.

The field function of UP in the momentum representation is defined
in exact analogy with the definition in \cite{14} except for the
above mentioned modification:
\begin{equation}\label{2.3}
\phi(q)=\int_{-\infty}^{+\infty}\phi(q,m^2)\,\omega(q,m^2)\,dm^2.
\end{equation}
The canonical commutation relations are not modified---they contain,
as in Ref.~\cite{14}, an additional delta-function
$\delta(m^2-m'{}^{2})$. Using the definition (\ref{2.3}) and the
commutation relations from \cite{14,15}, one get the propagator of a
scalar unstable field in spectral representation. Starting from the
standard definition of the Green's function, by straightforward
calculations we get the Lehmann-like spectral representation of the
propagator \cite{15,16}:
\begin{equation}\label{2.4}
D(q)=i\int dx\, exp\,(-iqx)\,\langle
0|\hat{T}\phi(x)\phi(0)|0\rangle
=i\int_{-\infty}^{+\infty}\,\frac{\rho(q,m^2)\,dm^2}{q^2-m^2+i\epsilon},
\end{equation}
where $\rho(q,m^2)=|\omega(q,m^2)|^2$. For the case $m^2<0$, that is
$m=i|m|$, we leave the physical range of the continuous (smearing)
mass. It should be noted, that the expression (\ref{2.4}) is
applicable for any values of $q^2$ and $m^2$ without restrictions.
Moreover, it is not directly connected with the expansion of the
Dyson type and perturbative constructions. So, the correct
definition of the function $\rho(q,m^2)$ makes it possible to escape
some problems which arise in a traditional approach.

Now, let us consider possible physical consequences of the presence
of negative mass parameter $m^2<0$ in the  spectral representation
(\ref{2.3}) and (\ref{2.4}). Negative component of the spectrum
leads to the states with imaginary mass parameters which usually
interpreted as tachyon states. The problem of the existence of
tachyons is under considerable discussion in the last decades. The
main attention is paid to the principal problems, such as a
violation of causality, tachyon vacuum, and radiation instability.
There is no unique and consistent quantum field theory of tachyons,
and various approaches are suggested to overcome the above mentioned
difficulties. We note, that the first problem relates to UP as an
observable object. Further we show that such object in the framework
of our effective model is described by the positive mass square
$M^2(q)$ and width $\Gamma(q)$. However, quantum field description
of UP as the continuous superposition with a tachyon component
encounters the problem of instability. In the third section, we
estimate the tachyon fraction for the case of fundamental UP. It is
found to be rather small; for instance, in the case of $Z$ boson it
is approximately a percent. The analysis of an expression for the
tachyon fraction leads to an interesting conclusion: tachyon
instability is intrinsic property of UP; it can be interpreted as
the cause of unstable particle decay.

The principal problem of the approach under consideration is to
define the spectral function $\rho(q,m^2)$ which is the main
characteristic of UP. Some phenomenological definitions were
considered in \cite{14}. Here, we consider the construction of this
function with an account of the most general formal and physical
arguments. First of all, we choose two-parametric distributions,
where the parameters are directly connected with the mean values of
mass and width of UP. These parameters are some functions of the
four-momentum, that is $M(q)$ and $\Gamma(q)$ are $q$-dependent
characteristics of UP. The probability density $\rho(q,m^2)$ has to
be normalized for any $q$. If $q\longrightarrow q_0$, where $q_0$ is
the threshold value of the momentum, the value
$\Gamma(q)\longrightarrow \Gamma(q_0)=0$. So, the particle become
stable, its mass go to a fixed value $M(q_0)=M_0$ and the function
$\rho(q,m^2)$ behaves as a delta-function $\delta (m-M_0)$. There
are three well-known approximations of the delta-function which are
continuously differentiable and normalized. These are as follows
\cite{22}:
\begin{equation}\label{2.5}
(a)\,\,\,\frac{\alpha}{\pi(1+\alpha^2
x^2)};\,\,\,(b)\,\,\,\frac{\alpha}{\sqrt{\pi}}\exp(-\alpha^2
x^2);\,\,\,(c)\,\,\frac{\alpha}{\pi}\frac{\sin \alpha x}{\alpha
x}\qquad (\alpha\to\infty).
\end{equation}
It is difficult to find the physical interpretation for the case of
the oscillating distribution $(c)$ in Eq.~(\ref{2.5}). As was
discussed in \cite{14}, the exponential distribution (b) can be
motivated when the smearing of mass is caused by stochastic
mechanism of UP interaction with vacuum. It was noted that fast
decreasing of the exponent does not describe some deeply virtual
processes. One can see that the distribution (a) is the well-known
Lorentzian distribution, where $\alpha \sim (\Gamma(q))^{-1}$ and
$x=m^2-M^2(q)$ or $x=m-M(q)$. It is easy to check that this function
is normalized to unity. So, the function (a) in Eq.~(\ref{2.5})
satisfies above mentioned formal and physical requirements and will
be used in derivation of the propagators of UP.

\section{\label{sec:3}Propagators of bosonic unstable particles}

In this section, we consider the structure of the propagators of
scalar and vector UP. From Eq.~(\ref{1.1}) and dimension requirement
we have to choose the parametrization $x=m^2-M^2(q)$ and
$\alpha=(q\Gamma(q))^{-1}$ or  $\alpha=(\Gamma(q))^{-2}$. Then, the probability density (\ref{2.5}a) takes the form
\begin{equation}\label{3.1}
\rho(q,m^2)=\frac{1}{\pi}\frac{\alpha^{-1}}{x^2+\alpha^{-2}}=
\frac{1}{\pi}\frac{q\Gamma(q)}{[m^2-M^2(q)]^2+q^2\Gamma^2(q)},
\end{equation}
where $M(q)$ and $\Gamma(q)$ are $q$-dependent parameters of the
distribution. According to Eqs.~(\ref{2.4}) and (\ref{3.1}) the
propagator of a scalar UP can be written as
\begin{equation}\label{3.2}
D(q)=\frac{1}{\pi}\int_{-\infty}^{+\infty}\frac{P(q)\,dm^2}
{(q^2-m^2+i\epsilon)\{[m^2-M^2(q)]^2+P^2(q)\}},
\end{equation}
where $P(q)=q\Gamma(q)$ and we omit $i$ for the simplicity of
consideration. The integral in (\ref{3.2}) can be calculated with the
help of the integration rule
\begin{equation}\label{3.3}
\int_{-\infty}^{+\infty}\frac{f(x)\,dx}{x\pm i\epsilon}=\mp i\pi
f(0)+ \mathcal{P}\int_{-\infty}^{+\infty}\frac{f(x)}{x}dx,
\end{equation}
which follows from the Sokhotski-Plemelj formula. In Eq.~(\ref{3.3})
$\mathcal{P}\int$ stand for the principal part of the integral. As a result
we have:
\begin{align}\label{3.4}
\Im D(q)={}&-\pi\rho(q,q^2)=\frac{-P(q)}{[q^2-M^2(q)]^2+P^2(q)};\notag\\
\Re
D(q)={}&\mathcal{P}\int_{-\infty}^{+\infty}\frac{\rho(q,m^2)\,dm^2}
{q^2-m^2}=\frac{q^2-M^2(q)}{[q^2-M^2(q)]^2+P^2(q)}.
\end{align}
It is easy to check that Eqs.~(\ref{3.4}) result in the following expression for the dressed propagator
of a scalar UP:
\begin{equation}\label{3.5}
D(q)=\frac{1}{q^2-M^2(q)+iP(q)},
\end{equation}
where $P(q)=q\Gamma(q)$. The same result can be got in a more simple
way with the help of contour integration. Let us rewrite the
function $D(q)$ as follows:
\begin{equation}\label{3.6}
D(q)=-\int_{-\infty}^{+\infty}\frac{\rho(q,m^2)\,dm^2}{m^2-(q^2+i\epsilon)}=-\frac{1}{\pi}
\int_{-\infty}^{+\infty}\frac{P(q)\,dm^2}{(m^2-z_0)(m^2-z_+)(m^2-z_-)},
\end{equation}
where $z_0=q^2+i\epsilon$ and $z_{\pm}=M^2(q)\pm iP(q)$. Analytical
continuation of the integrand function $f(z)$, where $m^2\to z$, has
three poles $z_0,z_+,z_-$ in the complex plane. It decreases as
$1/z^2$ for $z\to\infty$, that is, it satisfies the condition
$|f(z)|<M/|z|^{1+\delta}$ for $|z|>R_0$, where $M$ and $\delta$ are
positive numbers and $R_0\to\infty$. So, we can rearrange $D(q)$ as
follows:
\begin{equation}\label{3.7}
D(q)=\mp\frac{P(q)}{\pi}\oint_{C_{\pm}}\frac{dz}{(z-z_0)(z-z_+)(z-z_-)}
=2\pi i \sum_{k=z_0,z_+,z_-} \operatorname{Res}(f(z),z_k),
\end{equation}
where $\operatorname{Res}(f(z),z_k)$ is the residue at the pole $z_k$ and $C_{\pm}$ is a contour in the upper ($C_+$)
or lower ($C_-$) half of the complex $z$-plane. The simplest way to perform the integration is to go along the contour $C_-$ which contains only one pole
$z_-$:
\begin{align}\label{3.8}
D(q)={}&\frac{P(q)}{\pi}\oint_{C_-}\frac{dz}{(z-z_-)}\frac{1}{(z-z_+)(z-z_0)}\notag\\
={}&\frac{2i P(q)}{(z_--z_+)(z_--z_0)}=\frac{1}{q^2-M^2(q)+iP(q)}.
\end{align}
In Eqs.~(\ref{3.8}) we have used the equality $z_--z_+=-2iP(q)$. One
can check that the same result follows from the integration along
the contour $C_+$.

Thus, UP can be described at two different hierarchical
levels---``fundamental'' level by the spectral representation
(\ref{3.2}) and phenomenological one by the effective theory after
integrating out unobservable mass parameter $m^2$. In the framework
of this theory, UP is described by the observed physical values
$M^2(q)$ and $\Gamma(q)$, which can always be defined as a positive
quantity. So, at this phenomenological level UP has no explicit
tachyonic content that could lead to the problems noted in the
Introduction. At the first level, however, the approach under
consideration deals with explicitly tachyonic component in the field
function of UP. As it was noted earlier, it can lead to an
instability of the system. Let us estimate the tachyonic fraction,
that is the probability of the states with $m^2<0$. From the
Eq.(\ref{3.1}) it follows that this probability
$P_t(M(q),\Gamma(q))$ under the condition $M(q)/\Gamma(q)\ll 1$ is
as follows:
\begin{equation}\label{3.8a}
P_t(M(q),\Gamma(q))=\int_{-\infty}^0 \rho(m^2; M^2(q),
\Gamma(q))\,dm^2 \approx \frac{\Gamma(q)}{\pi M(q)}.
\end{equation}
Thus, for the majority of fundamental particles this value is very
small. However, for the case of vector boson ($Z$ and $W$) and
$t$-quark the tachyon fraction is appreciable ($~10^{-2}$). From the
Eq.(\ref{3.8a}) it follows that this fraction is defined by a value
of $\Gamma(q)/M(q)$, that is the value which quantifies the effects
of instability or finite-width effects (FWE) in the processes with
UP participation. This makes it possible to suggest the direct
connection between the tachyon component and instability of the
particle. The decay of UP can be caused, for instance, by the
instability of the tachyon vacuum.

To define the structure of the vector propagator, we assume that the
spectral function $\rho(q,m^2)$ is the same as for a scalar UP.
Using the standard vector propagator for a free vector particle with a
fixed mass, we get:
\begin{equation}\label{3.9}
D_{\mu\nu}(q)=\frac{1}{\pi}\int_{-\infty}^{+\infty}\frac{-g_{\mu\nu}+q_{\mu}q_{\nu}/(m^2-i\epsilon)}
{q^2-m^2+i\epsilon}\,\frac{P(q)\,dm^2}{[m^2-M^2(q)]^2+P^2(q)}.
\end{equation}
In the propagator term $q_{\mu}q_{\nu}/(m^2-i\epsilon)$ we use the
same rule of going around pole as in the denominator
$q^2-(m^2-i\epsilon)$. The integral in Eq.~(\ref{3.9}) can be evaluated
with the help of the formula (\ref{3.3}), however, it is easier to do
it using the method of contour integration. The integration along the lower contour $C_-$
gives:
\begin{align}\label{3.10}
D_{\mu\nu}(q)={}&-\frac{P(q)}{\pi}\oint_{C_-}\frac{(g_{\mu\nu}-q_{\mu}q_{\nu}/(z-i\epsilon))\,dz}
{(z-z_-)(z-z_+)(z-z_0)}\notag\\={}&-2iP(q)\frac{g_{\mu\nu}-q_{\mu}q_{\nu}/(z_-)}{(z_--z_+)(z_--z_0)}=
\frac{-g_{\mu\nu}+q_{\mu}q_{\nu}/(M^2(q)-iP(q))}{q^2-M^2(q)+iP(q)}\,.
\end{align}
One can check that the integration along the upper contour $C_+$ or
with the help of the formula (\ref{3.3}) leads to the same result.

We can see that both the scalar and vector propagators of UP can
be represented in the form with universal complex mass squared:
\begin{equation}\label{3.11}
D(q)=\frac{1}{q^2-M^2_P(q)};\qquad
D_{\mu\nu}(q)=\frac{-g_{\mu\nu}+q_{\mu}q_{\nu}/M^2_P(q)}{q^2-M^2_P(q)}\,,
\end{equation}
where $M^2_P(q)=M^2(q)-iP(q)$. Note that the dressed propagator of
a bosonic UP can be get from the ``free" propagator by the substitution $m^2-i\epsilon \longrightarrow
M^2(q)-iP(q)$. The inverse substitution,  $P(q)\to i0,
M^2(q)\to M^2_0$, leads to the ``free" propagator, if $q\to q_0$.

The $q$-dependent parameters $M(q)$ and $\Gamma(q)$ of the
propagator pole $s_R(q)=M^2_P(q)=M^2(q)-iq\Gamma(q)$ are a
generalization of the traditional pole definition of the mass and
width of UP, which is based on the complex-valued position of the
propagator pole $s_R-M^2_0-\Pi(s_R)=0$ \cite{1}. It should be noted
that such pole mass and width are defined in the non-physical region
of four-momentum squared $q^2=s_R$. In our analysis, propagator pole
is $q$-dependent, $s_R(q)=M^2(q)-iq\Gamma(q)$, so we have
$q$-dependent (running) pole definition of the mass and width. At
fixed point $q^2=M^2_{\rho}$ we get an ordinary definition
$s_R(M_{\rho})=M^2(M_{\rho})-M_{\rho}\Gamma(M_{\rho})$. The
correctness and gauge invariance of such definition are strongly
stipulated by the definition of the functions $M(q)$ and
$\Gamma(q)$.

The traditional pole definition of the mass and width follows from
the poles in the $S$-matrix. In practical calculations we deal with
an expansion of the $S$-matrix, which is defined by the interaction
Lagrangian. For instance, the process $e^+e^-\to Z\to f\bar{f}$ is
described at the tree level by the second order term and the pole in
the $S$-matrix in this approximation is defined by the pole of the
boson propagator. So, this pole depends on the scheme of inclusion
of the mass and width terms into the dressed propagator of $Z$-boson
(see the discussion in Refs.~\cite{23,24}). We consider $q^2$
dependent mass and width as the most general case. But, this scheme
is not obligatory for the consideration and we can include fixed
mass and width terms which is valid in the vicinity of the
resonance. As a result, we get a unique pole of the $S$-matrix and a
pole definition of the fixed mass and width.

\section{\label{sec:4}Propagator of a spinor unstable particle}

The propagator of a free fermion can be represented in two equivalent
forms:
\begin{equation}\label{4.1}
\hat{D}(q)=\frac{1}{\hat{q}-m+i\epsilon}=\frac{\hat{q}+m-i\epsilon}{q^2-(m-i\epsilon)^2}.
\end{equation}
According to the heuristic rule for constructing the dressed propagator, we have to make the substitution $m\to M(q)$ and $i\epsilon\to
i\Gamma(q)$, where the dimension $[\epsilon]=[m]$ was taken
into account. Then, the dressed propagator of the spinor UP takes
the form
\begin{equation}\label{4.2}
\hat{D}(q)=\frac{\hat{q}+M(q)-i\Gamma(q)}{q^2-(M(q)-i\Gamma(q))^2}=\frac{\hat{q}+M_P(q)}{q^2-M^2_P(q)},
\end{equation}
where $M_P(q)=M(q)-i\Gamma(q)$ is $q$-dependent pole-type complex
mass. Now, we show that the expression (\ref{4.2}) can be derived in a
more systematic way with the help of the spectral representation:
\begin{equation}\label{4.3}
\hat{D}(q)=\int\frac{\hat{q}+m}{q^2-(m-i\epsilon)^2}\,\rho(q,m)\,dm\,,
\end{equation}
where the integration range is not defined yet. The spectral
function $\rho(q,m)$ for fermions differs from the bosonic one,
because of another parametrization $M(q)=M_0+\Re\Sigma(q)$ and
$\Gamma(q)=\Im\Sigma(q)$. So, we have to take $x=m-M(q)$ and
$\alpha=\Gamma(q)$ in the general expression for $\rho(q,m)$ in
Eq.~(\ref{2.5}a). As a result, the spectral
function  for the case of the spinor UP is as follows:
\begin{equation}\label{4.4}
\rho(q,m)=\frac{1}{\pi}\,\,\frac{\Gamma(q)}{[m-M(q)]^2+\Gamma^2(q)}
=\frac{1}{\pi}\frac{\Gamma(q)}{(m-M_-(q))(m-M_+(q))},
\end{equation}
where $M_{\pm}(q)=M(q)\pm \Gamma(q)$. The main difference between
boson and spinor cases is a presence of the linear term $m$ instead
the quadratic one $m^2$, which is defined at the whole real axis
$m^2 \in (-\infty,+\infty)$. Here, we restrict the problem by the
formal definition and consider a straightforward relation between
the bosonic parameter range and spinor one. Thus, we have two
intervals $(\pm i\infty , i0; 0 , \infty)$ for the value $m$. In the
method of contour integration the signs $\pm$ correspond to
integration along the contours $C_{\pm}$, which enclose the first or
fourth quadrants of the complex plane. Then, from Eqs.~(\ref{4.3})
and (\ref{4.4}) it follows:
\begin{equation}\label{4.5}
\hat{D}_{\pm}(q)=\pm\frac{\Gamma(q)}{\pi}\int_{C_{\pm}}\frac{(\hat{q}+z)\,dz}
{(z^2-z^2_0)(z-z_-)(z-z_+)}\,,
\end{equation}
where $z^2_0=q^2+i\epsilon$, $z_{\pm}=M_{\pm}(q)$ and $C_{\pm}$ are
the above described contours. By simple and straightforward
calculations one can see that the result (\ref{4.2}) follows from
the integration along the contour $C_-$, while the integration along
the $C_+$ leads to non-physical result. This is likely caused by the
presence of the branch point $z^2_0$ in the first quadrant. From
Eq.~(\ref{4.5}) it follows:
\begin{align}\label{4.6}
\hat{D}_{-}(q)=&-\frac{\Gamma(q)}{\pi}\int_{C_{-}}\frac{dz}{z-z_-}\,\frac{\hat{q}+z}
{(z^2-z^2_0)(z-z_+)}\notag\\=&-2i\Gamma(q)\frac{\hat{q}+z_-}{(z^2_--z^2_0)(z_--z_+)}
=\frac{\hat{q}+M_P(q)}{q^2-M^2_P(q)}.
\end{align}
The integration along the contour $C_+$ leads to a more complicated
expression. Note, that while the spinor $q$-dependent complex mass
$M_P(q)=M(q)-i\Gamma(q)$ differs from the bosonic one, it has,
however, the same pole-type structure. Then, the pole definition of
the mass and width of the spinor UP is
$M_P(M_{\rho})=M(M_{\rho})-i\Gamma(M_{\rho})$.

\section{\label{sec:5}Conclusion}
The definitions of the mass and width of UP, as a rule, are closely
connected with the construction of the dressed propagators. There
are two main definitions, which follows from the on-shell
re-normalization and pole structure of the propagators (see
Introduction). We considered the general structure
of the propagators of UP in the phenomenological approach based on the
spectral representation. The spectral function was defined with
account of the some formal and physical suggestions which naturally
arise in the model of UP with continuous mass.

In this work, we analyzed a specific case---the spectral function
depends on $q$-dependent parameters and random mass variable $m^2$
defined in the interval $(-\infty,+\infty)$. So, the variable $m$
can be imaginary, which is beyond the physical domain in the model
of UP with continuous (smeared) mass \cite{14}. It was shown that
this leads to the $q$-dependent complex mass which gives the pole
definition of the mass and width of UP. We also suggest that the
imaginary component of mass spectrum, which corresponds to tachyonic
states, causes the instability of particle.

We have proved a simple heuristic rule for constructing the dressed
propagators, which is based on the definition of $q$-dependent mass
and width of UP. Physical $q$-dependence of the UP width makes it
possible to introduce naturally the stable particle limit near
threshold momentum and the narrow width approximation
$\Gamma(q^2)\to 0$ when $q^2\to q^2_0$. Note, however, that the
physical nature of the random mass parameter $m^2$ is vague at
$m^2<0$.

\end{document}